\documentclass[aps,pre,twocolumn]{revtex4}
\usepackage{graphicx}

\begin{document}

\title{Resonant Analytic Fields Applied to Generic Multi-state Systems}

\author{Toshiya Takami}
\affiliation{Research Institute for Information Technology, Kyushu University,
Fukuoka 812-8581, Japan}
\author{Hiroshi Fujisaki}
\affiliation{Institut f\"ur Physikalische und Theoretische Chemie,
J.~W.~Goethe-Universit\"at, Max-von-Laue-Str. 7,
D-60438 Frankfurt am Main, Germany}

\date{\today}

\begin{abstract}
Rotating-wave approximation and its validity in multi-state quantum systems
are studied through analytic approach.
Their applicability is also verified from the viewpoint of generic states
by the use of direct numerical integrations of the Schr\"odinger equation.
First, we introduce an extension of the rotating-wave
approximation for multi-state systems.
Under an assumption that a smooth transition is induced by the optimal field,
we obtain three types of analytic control fields
and demonstrate their validity and deficiency
for generic systems represented by random matrices.
Through the comparison,
we conclude that the analytic field based on our coarse-grained approach
outperforms the other ones for generic quantum systems
with a large number of states.
Finally, the further extension of the analytic field is introduced
for realistic chaotic systems and its validity is shown
in banded random matrix systems.
\end{abstract}


\maketitle

\section{Introduction}

Theoretical and experimental studies of controlling quantum states
have been attracting much attention because of the theoretical
progress in the field of quantum computing \cite{NC00} and of the
technical developments in manipulating atomic and molecular systems
\cite{RZ00}.  Various control schemes are known in these fields: A
$\pi$ pulse in a transition between two eigenstates \cite{AE87}, the
nonadiabatic transitions induced by laser fields \cite{TN98}, the
STIRAP scheme by a counterintuitive pulse sequence for more than
three-level systems \cite{BTS98}, etc.  If we use an electronically
excited state in the controlled system, a simple control of a
pulse-timing can selectively break a chemical bond by pump and dump
pulses \cite{TR85}.  When two pathways exists from an initial state to
a target state, quantum mechanical interference can be utilized to
modify the ratio of products, which is called a coherent control
scheme \cite{SB03}.

These control schemes are effective for a certain class of processes
but are not versatile for general multilevel-multilevel transitions.
When we consider to design quantum devices with a large number of
states interacting with a complex environment or with short-time laser
pulses, such a multi-state control problem should be considered.
Optimal control theory \cite{PDR88,ZBR98} and genetic algorithms
\cite{RHR04,Miller06} are most successful methods to solve this kind
of complicated problems, but its implementation and interpretation can
be still difficult: What kinds of dynamical processes are involved in
the controlled dynamics?

On the other hand, we know that highly excited systems exhibit quantum
chaotic features \cite{Gutzwiller90}.  Such a "complex" quantum system
driven by an external field \cite{Igarashi06,SK2008} is modeled
by fully random matrix systems
with a parameter, where statistical properties of eigenenergies and
eigenvectors \cite{GRMN90,TKM91,TH92,ZD93} are characterized under
universality classes \cite{Haake01}.  These statistical properties
stem from multilevel-multilevel interactions of eigenstates, which are
related to the existence of many avoided crossings \cite{TKM92,ZD93a}.
Hence it is necessary to consider the interaction between many
eigenstates when we study dynamics in such a system.  Gong and Brumer
applied the coherent control method \cite{SB86,SB03} to a quantum
chaos system \cite{GB01-PRL,GB05} and its prediction has been recently
confirmed by experiment.

Several attempts have also been done to obtain control fields
analytically in multi-level systems \cite{AWS98,KAS02}.  We also
derived an analytic optimal field to control the fully random matrix
systems \cite{TF07}, which is based on the naive idea of
``coarse-grained approach'' \cite{TFM05,TF04}, and the results are
promising.  In this paper, at first, we introduce several analytic schemes
from the viewpoint of applicability to generic quantum systems, and
confirm the so-called rotating-wave approximation for multistate systems.
Next, we improve our previous approach \cite{TF07} to deal with
more realistic quantum systems with a banded random Hamiltonian, since
the most realistic quantum systems may be modeled with such a banded
random matrix \cite{Haake01}.  The validity of the extended analytic
field is evaluated by numerically solving the Schr\"odinger equation
for the multilevel-multilevel control problem.

\section{Analytic Fields for Multi-state Systems}

In a simple two-state transition problem,
the Bloch vector in a three dimensional space
is introduced to represent the transition dynamics \cite{AE87}.
The rotating-wave approximation (RWA) is also
introduced if we note slowly varying dynamics
on a rotating frame in the Bloch space,
where a transition by a $\pi$-pulse
is also represented as a rotation with an angle $\pi$.
The first problem considered in this section is
``what is the natural extension
of this representation in multi-state quantum systems.''

\subsection{The rotating-wave approximation}
\label{sec:rwa}

Actual procedure of the RWA for two-state problems is to
ignore off-resonant terms in the Schr\"odiner equation,
where an intuitive interpretation of this is
that a rapid motion induced by oscillating terms can be averaged out
and only slow dynamics by near-resonant terms remain.
We can extend this idea to the multi-state dynamics
induced by an external field.

We consider the Hamiltonian system
\begin{equation}
\label{eq:Hamiltonian}
  \hat H(t)=\hat H_0+\varepsilon(t)\hat V
\end{equation}
with an interaction term $\varepsilon(t)\hat V$ by an external field
$\varepsilon(t)$.
If we introduce an eigenstate representation
\begin{equation}
\label{eq:es-rep}
  \left|\psi(t)\right\rangle
  =\sum_ja_j(t)|\varphi_j\rangle\exp\left[\frac{E_jt}{i\hbar}\right],
\end{equation}
the time evolution of $\{a_j(t)\}$ is represented by
\begin{equation}
  \dot a_k(t)=\frac{\varepsilon(t)}{i\hbar}\sum_jV_{kj}a_j(t)
   \exp\left[\frac{(E_j-E_k)t}{i\hbar}\right],
\end{equation}
where $E_j$ and $|\varphi_j\rangle$ are
the $j$-th eigenvalue and eigenstate, respectively,
and $V_{kj}$ is a shorthand of $\langle\varphi_k|\hat V|\varphi_j\rangle$.

We consider the case \cite{KAS02} that the external field
is a sum of oscillating terms
with frequencies of the transition energies $\omega_{jk}\equiv(E_j-E_k)/\hbar$,
\begin{equation}
\label{eq:field-restriction}
  \varepsilon(t)
  =\sum_{j,k\ (\ne j)}\varepsilon_{jk}\ e^{-i\omega_{jk}t}+{\mathrm c.c},
  \quad\omega_{jk}\equiv\frac{E_j-E_k}{\hbar}.
\end{equation}
According to the standard procedure of the rotating-wave approximation,
we can separate the oscillating terms in the Schr\"odinger equation
into slowly varying ones and rapidly changing ones.
Further, we can introduce an approximation that all the oscillating terms
are ignored when we consider the limit of infinitely long transition time
under a non-degenerate condition between energy differences,
\begin{equation}
  E_j-E_k\ne E_{j'}-E_{k'}\quad\hbox{for $j\ne j'$ and $k\ne k'$}.
\end{equation}
Note that the validity of this simplification of the Schr\"odinger
equation is not trivial although this seems a direct extension
of the usual RWA.

\subsection{Slow transition dynamics}
\label{sec:slowDynamics}

Consider a control problem from an initial state $|\Phi_0\rangle$
at $t=0$ to a target state $|\Phi_T\rangle$ at $t=T$.
The interaction representation \cite{J.J.Sakurai94} is introduced by
\begin{equation}
  |\phi_0(t)\rangle=\hat U_0(t,0)|\Phi_0\rangle,\quad
  |\chi_0(t)\rangle=\hat U_0(t,T)|\Phi_T\rangle,
\end{equation}
where $\hat U_0(t_2,t_1)$ is a propagator by $\hat H_0$
from $t=t_1$ to $t=t_2$.

An overlap between $|\phi_0(t)\rangle$ and $|\chi_0(t)\rangle$ is
parameterized by an angle $\Theta$ ($0\le\Theta<\pi/2$) with
a phase $\alpha$ ($0\le\alpha<2\pi$),
\begin{equation}
\label{eq:innerProduct}
  \langle\phi_0(t)|\chi_0(t)\rangle
  =\langle\Phi_0|\hat U_0(0,T)|\Phi_T\rangle
  =ie^{i\alpha}\sin\Theta
\end{equation}
which includes an orthogonal case ($\Theta=0$).
For the case of $\Theta\ne0$, the phase $\alpha$ is uniquely determined.
Then, we can introduce orthogonal basis states,
\begin{equation}
  |\tilde\phi_0(t)\rangle=|\phi_0(t)\rangle,\quad
  |\tilde\chi_0(t)\rangle
    =\frac{|\chi_0(t)\rangle-ie^{i\alpha}\sin\Theta|\phi_0(t)\rangle}
          {\cos\Theta}.
\end{equation}

From the result of optimally controlled dynamics
in random matrix systems \cite{TF07},
the time-evolution is assumed to be
\begin{equation}
\label{eq:smooth-dynamics}
  |\psi(t)\rangle=|\tilde\phi_0(t)\rangle\cos\theta
    -ie^{-i\alpha}|\tilde\chi_0(t)\rangle\sin\theta,
\end{equation}
in the limit of $T\rightarrow\infty$.
Substituting it into the Schr\"odinger equation,
we obtain
\begin{widetext}
\begin{equation}
\label{eq:restricted-equation}
  \dot\theta\left[
    -|\tilde\phi_0(t)\rangle\sin\theta
    -ie^{-i\alpha}|\tilde\chi_0(t)\rangle\cos\theta
  \right]
  =\frac{\varepsilon(t)}{i\hbar}\hat V\left[
    |\tilde\phi_0(t)\rangle\cos\theta
    -ie^{-i\alpha}|\tilde\chi_0(t)\rangle\sin\theta
  \right].
\end{equation}
\end{widetext}
If we operate $\langle\tilde\phi_0(t)|$ from the left, the relation
\begin{equation}
\label{eq:theta-dot1}
  \dot\theta
  =-\frac{\varepsilon(t)}{i\hbar}\left[
    \langle\tilde\phi_0(t)|\hat V|\tilde\phi_0(t)\rangle\cot\theta
   -ie^{-i\alpha}\langle\tilde\phi_0(t)|\hat V|\tilde\chi_0(t)\rangle
  \right]
\end{equation}
is obtained, and the operation of $\langle\tilde\chi_0(t)|$
gives another relation
\begin{equation}
\label{eq:theta-dot2}
  \dot\theta
  =\frac{\varepsilon(t)}{i\hbar}\left[
    ie^{i\alpha}\langle\tilde\chi_0(t)|\hat V|\tilde\phi_0(t)\rangle
   +\langle\tilde\chi_0(t)|\hat V|\tilde\chi_0(t)\rangle\tan\theta
  \right].
\end{equation}
It is almost impossible to satisfy these equations strictly
since such a field realizing the given dynamics Eq.(\ref{eq:smooth-dynamics})
does not exist always.
However, approximate fields can be obtained
within a restriction to the form of Eq.(\ref{eq:field-restriction}).
In the following, we try to determine $\varepsilon(t)$ under the RWA
for several special cases.

\subsubsection{Exact field for a transition from an eigenstate}
\label{sec:exact-field}

The simplest example \cite{KAS02} is given by the case that
$|\Phi_0\rangle$ is an eigenstate,
and $|\Phi_T\rangle$ is a linear combination of multiple eigenstates,
\begin{equation}
  |\Phi_0\rangle=|\varphi_0\rangle,\quad
  |\Phi_T\rangle
     =\sin\Theta|\varphi_0\rangle+\cos\Theta\sum_{j\ne0}d_j|\varphi_j\rangle.
\end{equation}
The angle $\Theta$ satisfies $0\le\Theta<\pi/2$,
and the normalization 
\begin{equation}
  \sum_{j\ne0}|d_j|^2=1
\end{equation}
should be satisfied.
We introduce
\begin{eqnarray}
  |\phi_0(t)\rangle
  &=&|\varphi_0\rangle e^{E_0t/i\hbar},\\
  |\chi_0(t)\rangle
  &=&\sin\Theta|\varphi_0\rangle e^{E_0(t-T)/i\hbar}\nonumber\\
  &&+\cos\Theta\sum_{j\ne0}d_j|\varphi_j\rangle e^{E_j(t-T)/i\hbar},
\end{eqnarray}
and, from the relation Eq.(\ref{eq:innerProduct}),
the phase 
is determined by
\begin{equation}
  \alpha=\frac{E_0T}{\hbar}-\frac{\pi}2.
\end{equation}
Then, the orthogonal basis states are
\begin{equation}
  |\tilde\phi_0(t)\rangle=|\varphi_0\rangle e^{E_0t/i\hbar},\quad
  |\tilde\chi_0(t)\rangle
  =\sum_{j\ne0}d_j|\varphi_j\rangle e^{E_j(t-T)/i\hbar}.
\end{equation}
The multiplication of $\langle\varphi_0|$ from the left to
Eq.(\ref{eq:restricted-equation}) gives a relation
\begin{equation}
\label{eq:s2m1}
  \dot\theta=-\frac{\varepsilon(t)}{i\hbar}\left[
    V_{00}\cot\theta+\sum_{j\ne0}V_{0j}d_j\ e^{-i\omega_{j0}(t-T)}
  \right],
\end{equation}
and the operation of $\langle\varphi_j|$ ($j\ne0$) from the left
gives another relation
\begin{equation}
\label{eq:s2m2}
  \dot\theta\ d_j=\frac{\varepsilon(t)}{i\hbar}\left[
    V_{j0}\ e^{i\omega_{j0}(t-T)}
   +\sum_kV_{jk}d_k\ e^{i\omega_{jk}(t-T)}\tan\theta
  \right].
\end{equation}
If we restrict $\varepsilon(t)$ to a sum of terms with transition frequencies
from the $0$-th eigenstate to the $j$-th eigenstate, i.e.,
\begin{equation}
\label{eq:Er-field}
  \varepsilon(t)
  =\sum_{j\ne0}\varepsilon_{j0}\ e^{-i\omega_{j0}(t-T)}+{\mathrm c.c},
\end{equation}
we obtain the optimal field
\begin{equation}
\label{eq:E-field}
  \varepsilon_{\rm e}(t)
  =i\hbar\Omega\sum_{j\ne0}\frac{d_j}{V_{j0}}e^{-i\omega_{j0}(t-T)}
   +\hbox{c.c.}
\end{equation}
by ignoring all the oscillating terms
(for details, see Appendix \ref{apx:E-field}),
where $\Omega$ ($>0$) is a constant or a slowly varying function of time.
If we substitute $\varepsilon_{\rm e}(t)$ into Eq.(\ref{eq:s2m1})
or Eq.(\ref{eq:s2m2}), $\dot\theta=\Omega$ is easily shown.
Since the target state $|\Phi_T\rangle$ is realized by
$\theta=\frac{\pi}{2}$ at $t=T$ in Eq.(\ref{eq:smooth-dynamics}),
the smallest rotation angle is $\frac{\pi}{2}-\Theta$.
Thus, the smallest $\Omega$ is determined by
\begin{equation}
  \Omega=\left.\left(\frac{\pi}{2}-\Theta\right)\right/T,
\end{equation}
which induces the perfect control $|\langle\Phi_T|\psi(t)\rangle|=1$ at
the target time $t=T$.

We note that $\varepsilon_{\rm e}(t)$ is obtained under the assumption
of a slow transition Eq.(\ref{eq:smooth-dynamics})
with a sufficiently long target time $T$.
Then, the quantum state $|\psi(t)\rangle$ driven
by $\varepsilon_{\rm e}(t)$ stays in a plane determined by the two states,
$|\tilde\phi_0(t)\rangle$ and $|\tilde\chi_0(t)\rangle$,
during the controlled dynamics from $t=0$ to $t=T$.

\subsubsection{Approximate field for a transition between multi-level states}
\label{sec:approximate-field}

We study the case that the initial state $|\Phi_0\rangle$ and
the target state $|\Phi_T\rangle$ are two different linear combinations of
many eigenstates
\begin{equation}
  |\Phi_0\rangle=\sum_jc_j|\varphi_j\rangle,\qquad
  |\Phi_T\rangle=\sum_jd_j|\varphi_j\rangle.
\end{equation}
If we use Eq.(\ref{eq:innerProduct}) for the phase,
the pair of orthogonal states is defined by
\begin{widetext}
\begin{eqnarray}
\label{eq:state-phi}
  |\tilde\phi_0(t)\rangle&=&\sum_jc_j|\varphi_j\rangle e^{E_jt/i\hbar}\\
\label{eq:state-chi}
  |\tilde\chi_0(t)\rangle
    &=&\sum_j\frac{d_j\ e^{-E_jT/i\hbar}-ie^{i\alpha}c_j\sin\Theta}{\cos\Theta}
       |\varphi_j\rangle e^{E_jt/i\hbar}
    \equiv\sum_j\tilde d_j|\varphi_j\rangle e^{E_jt/i\hbar}.
\end{eqnarray}
\end{widetext}
Based on the assumption that the controlled state
represents a smooth rotation Eq.(\ref{eq:smooth-dynamics})
and that $\varepsilon(t)$ has a restricted form
Eq.(\ref{eq:field-restriction}),
we can define
\begin{equation}
\label{eq:A-field}
  \varepsilon_{\rm a}(t)=\hbar\Omega\sum_{j,k\ (\ne j)}\left[
    \frac{c_j\tilde d_k^*}{V_{jk}}e^{i\alpha}e^{-i\omega_{jk}t}
   +\hbox{c.c.}
  \right]
\end{equation}
as an approximate control field (for details, see Appendix \ref{apx:A-field}).
If we substitute $\varepsilon_{\rm a}(t)$ into Eq.(\ref{eq:theta-dot1})
and Eq.(\ref{eq:theta-dot2}), we obtain
\begin{equation}
\label{eq:theta-dot}
  \dot\theta=\Omega\left[1+O(N^{-1})\right]
\end{equation}
by ignoring all the oscillating terms,
where $N$ is the number of eigenstates contained in the state,
and $\{c_j\}$ and $\{\tilde d_k\}$ are assumed as random complex numbers.
Note that the approximate field $\varepsilon_{\rm a}(t)$ is
valid in the limit of $N\rightarrow\infty$ as well as $T\rightarrow\infty$.

\subsubsection{Another field for a transition containing many eigenstates}
\label{sec:another-field}

We use another approach to the approximate analytic field
in the limit of infinitely many eigenstates.
By the assistance of the optimal control theory \cite{TF07},
we can introduce a control field defined by
\begin{equation}
\label{eq:CG-field}
  \varepsilon_{\rm cg}(t)
  =\frac{2\hbar\Omega}{\left|V\right|^2}{\mathrm Re}\left[
    e^{-i\alpha}\langle\tilde\phi_0(t)|\hat V|\tilde\chi_0(t)\rangle
  \right]
\end{equation}
where
\begin{equation}
  \left|V\right|^2\equiv\frac1T\int_0^T\left|
    \langle\tilde\phi_0(t)|{\hat V}^2|\tilde\chi_0(t)\rangle
  \right|^2dt.
\end{equation}
Substituting $\varepsilon_{\rm cg}(t)$ into
Eqs. (\ref{eq:theta-dot1}) and (\ref{eq:theta-dot2}),
we obtain
\begin{equation}
  \theta(T)=\int_0^T\dot\theta(t)dt
  \approx\Omega T
\end{equation}
under the limit of $T\rightarrow\infty$ and $N\rightarrow\infty$.
This is the optimal field based on the coarse-grained approach
we derived before \cite{TFM05,TF04}.

\section{Numerical Evaluation for Generic Systems}
\label{sec:numerical}

In the previous section,
we have obtained several optimal fields for multi-state transitions, i.e.,
$\varepsilon_{\rm e}(t)$, $\varepsilon_{\rm a}(t)$,
and $\varepsilon_{\rm cg}(t)$.
Applicability of these fields depends mainly on the validity of the RWA.
Furthermore, for the cases of $\varepsilon_{\rm e}(t)$ and
$\varepsilon_{\rm a}(t)$, 
the matrix element $V_{jk}$ in the denominator may deteriorate the results
while the effect of the number of states should be checked
for $\varepsilon_{\rm a}(t)$ and $\varepsilon_{\rm cg}(t)$.
All these points can be verified in a direct numerical integration
of the Schr\"odinger equation for generic quantum systems
with a random matrix Hamiltonian under Gaussian Orthogonal Ensemble (GOE)
or Gaussian Unitary Ensemble (GUE) \cite{Haake01}.

The GOE random matrix represents the case of ``complex'' Hamiltonian systems
with a time-reversal symmetry \cite{Haake01}.
This is constructed by a real symmetric matrix
with off-diagonal elements $v$ subject to a distribution function
\begin{equation}
\label{eq:Pgoe}
  P_{\rm goe}(v)dv\propto\exp\left[-\frac{v^2}{2\Delta^2}\right]dv.
\end{equation}
This means that the density of $v$ has a maximum value at $v=0$.

The GUE random matrix represents Hamiltonian systems
without time-reversal symmetry \cite{Haake01}.
This is constructed by a Hermitian matrix
with off-diagonal elements $z=x+iy$ subject to a distribution function
\begin{equation}
  P_{\rm gue}(z)dz\propto\exp\left[-\frac{|z|^2}{2\Delta^2}\right]dz
\end{equation}
If we take a polar-angle representation ($r$, $\phi$),
$z=x+iy=re^{i\phi}$, the distribution function for $r$ is
\begin{equation}
\label{eq:Pgue}
  P_{\rm gue}(r)dr\propto r\ \exp\left[-\frac{r^2}{2\Delta^2}\right]dr
\end{equation}
This means that the density of $r$ vanishes at $r=0$.
Thus, the number of small off-diagonal elements in GUE matrices
are relatively fewer than that in GOE matrices.

\subsection{Final overlaps of transition from an eigenstate}

\begin{figure}
\begin{center}
\includegraphics[scale=0.55]{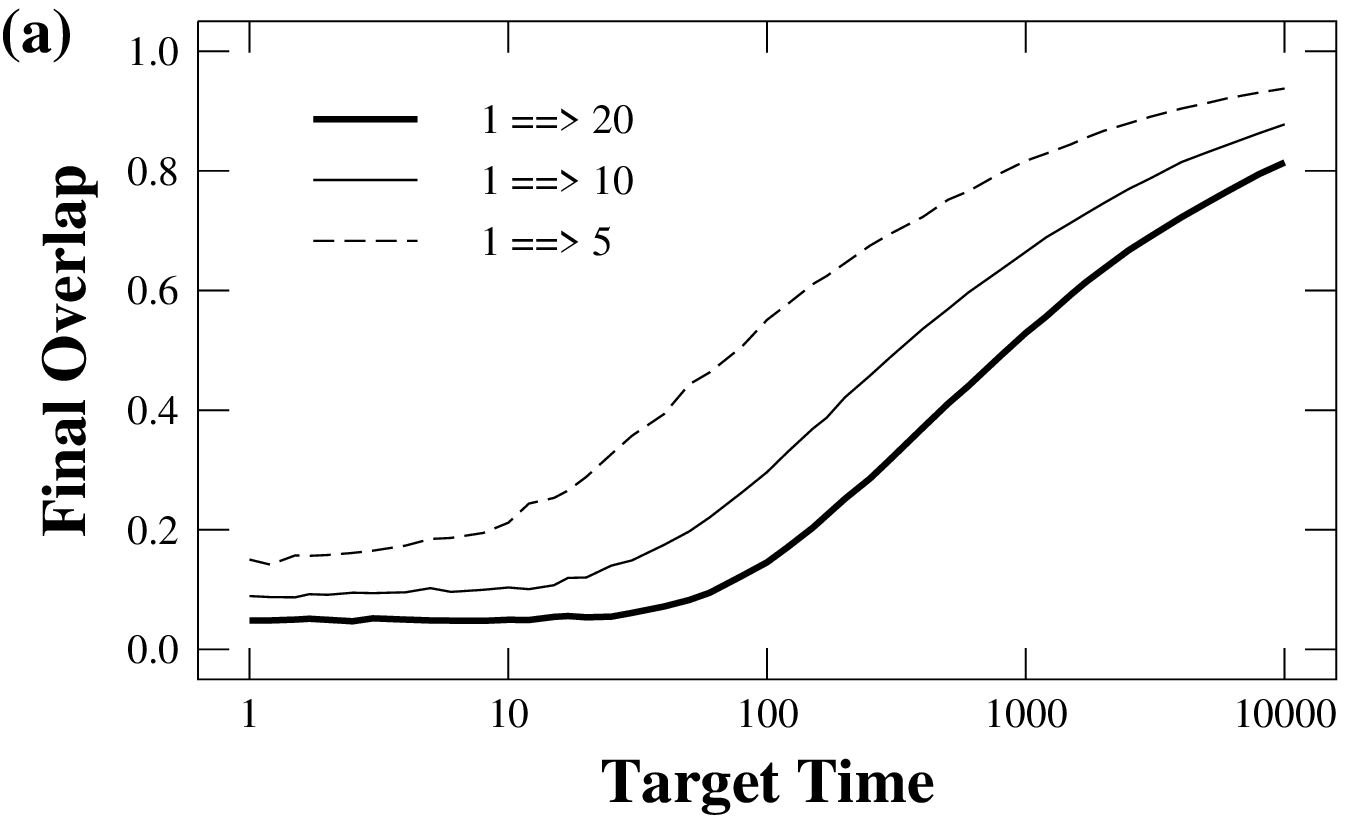}\\
\includegraphics[scale=0.55]{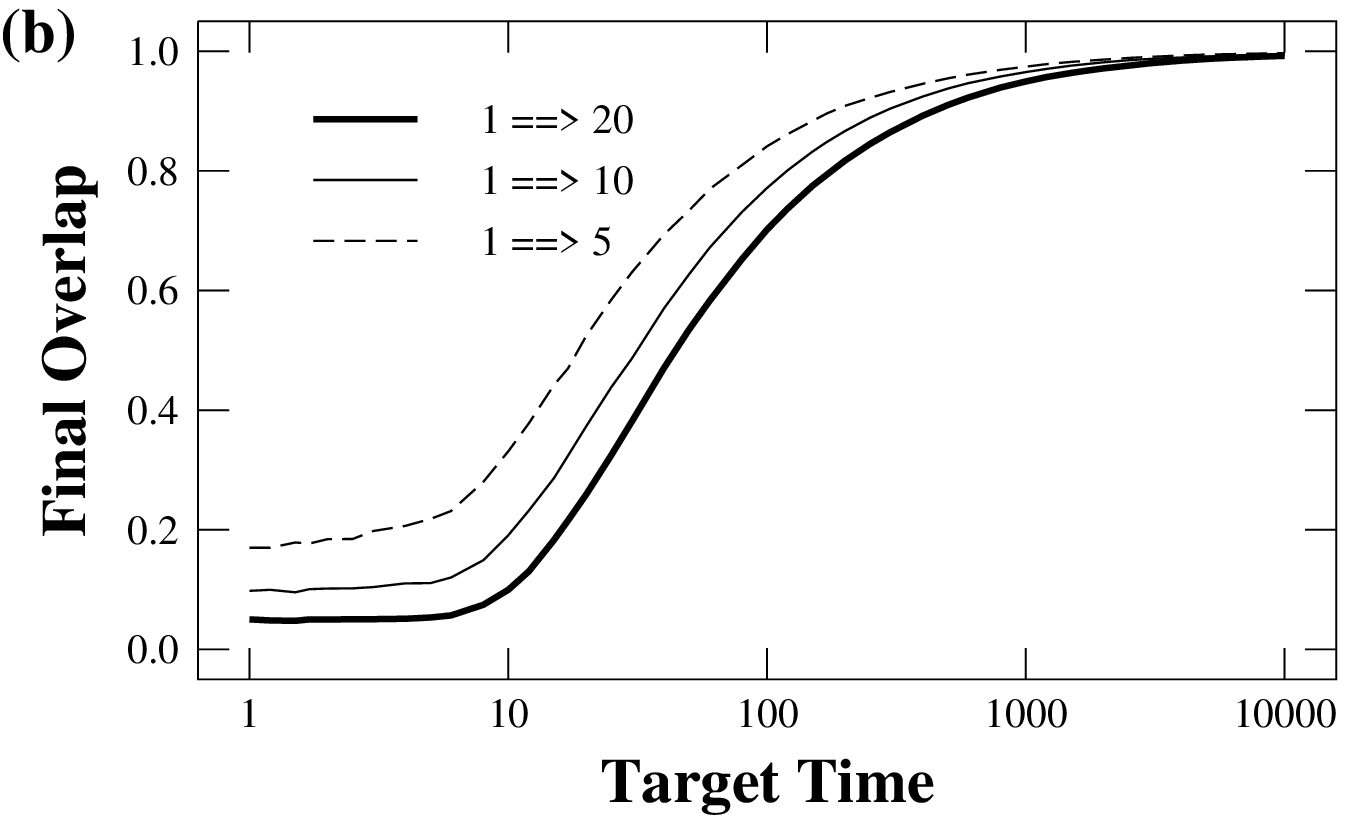}
\caption{Final overlaps by $\varepsilon_{\rm e}(t)$ for the transition
from an eigenstate to generic states constructed from 5 (dashed curve),
10 (thin curve), and 20 eigenstates (thick curve)
in (a) GOE and (b) GUE random matrix systems.
Each curve is obtained as an average over 100 different random configurations
of $H_0$, $V$, and $\{d_j\}$.}
\label{fig:o2m}
\end{center}
\end{figure}

For the numerical evaluation,
Hamiltonian matrices $\hat H_0$ and $\hat V$ are created
by random numbers subject to GOE or GUE.
$\hat H_0$ is scaled so that
the average spacing $\Delta E\equiv\langle|E_j-E_{j-1}|\rangle$ is unity,
and the target time $T$ is shown in units of $\hbar/\Delta E$.
The exact control field $\varepsilon_{\rm e}(t)$ is defined
immediately after the target state $|\Phi_T\rangle$,
i.e., coefficients $\{d_j\}$, is determined.
Then, the final overlap $|\langle\Phi_T|\psi(T)\rangle|$ is
obtained by numerical integration of the controlled dynamics by
$\varepsilon_{\rm e}(t)$ written in the Schr\"odinger equation
without RWA.
Fig.\ref{fig:o2m} represents the result obtained as an ensemble average
over 100 different samples of $H_0$, $V$, and $\{d_j\}$.

It is shown that
$\varepsilon_{\rm e}(t)$ works only for sufficiently large $T$
in spite of the exact control field under the RWA,
which means that the RWA is valid only for such a large $T$.
The reason why the RWA breaks down for smaller $T$
is that the field amplitude for the frequency
corresponding to the $0\rightarrow j$ transition becomes
accidentally large when $|V_{j0}|\approx0$.
In order to avoid the breakdown,
we must use the field $\varepsilon_{\rm e}(t)$ with the smaller amplitude
which is realized by a larger $T$.
If we take the limit $T\rightarrow\infty$, in this case,
it is expected that $|\langle\Phi_T|\psi(T)\rangle|\rightarrow1$
since $\varepsilon_{\rm e}(t)$ is the exact control field
within the RWA.

Difference in the performance of this field
with respect to the universality class
is also explained by the difference in
probability distribution functions for smaller $|V_{j0}|$
(See Eq.(\ref{eq:Pgoe}) and Eq.(\ref{eq:Pgue})).
When we use larger matrices, larger target times will be necessary
to obtain a certain final overlap since the probability for
accidental divergence of the amplitude tends to occur
for many off-diagonal elements.

\subsection{Final overlaps in transition between multi-level states}

\begin{figure}
\begin{center}
\includegraphics[scale=0.55]{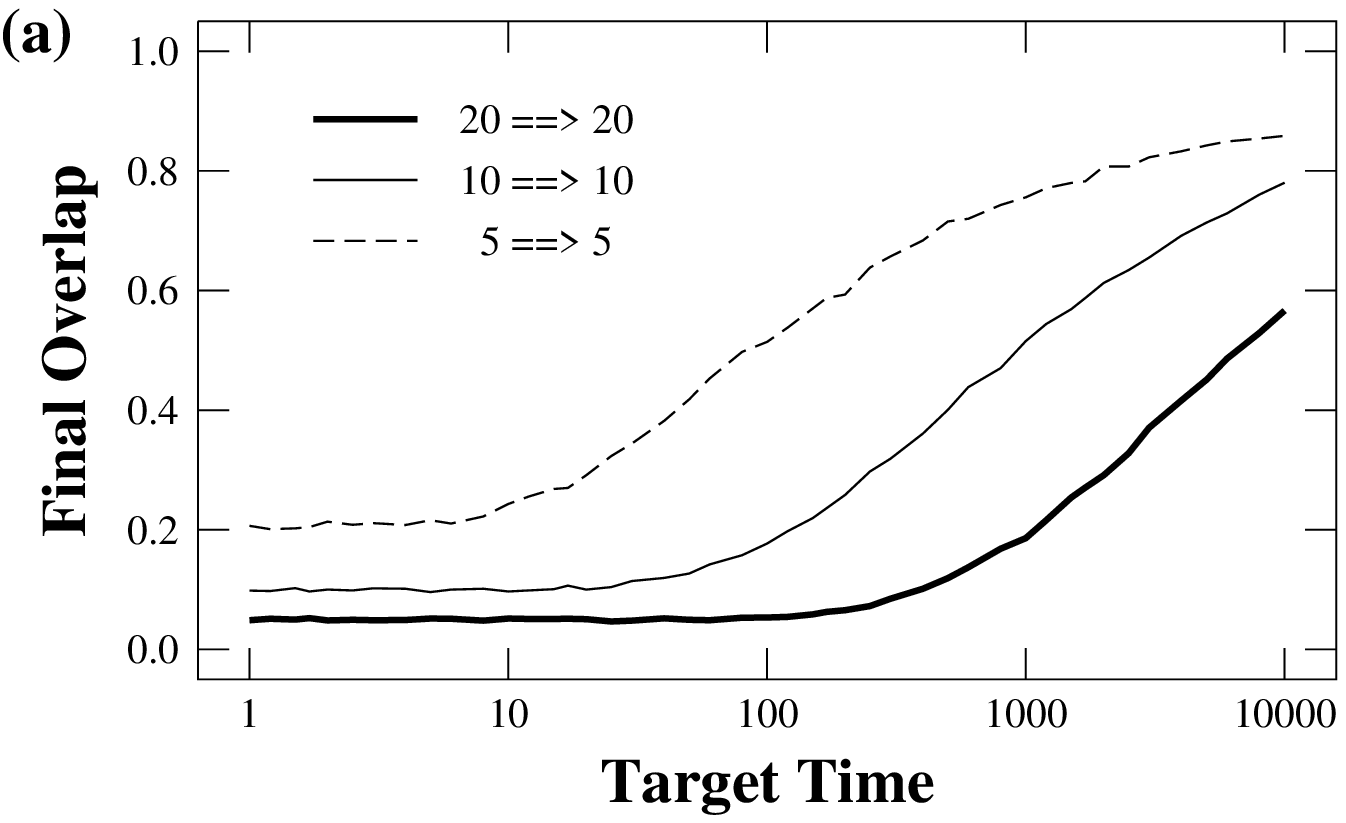}\\
\includegraphics[scale=0.55]{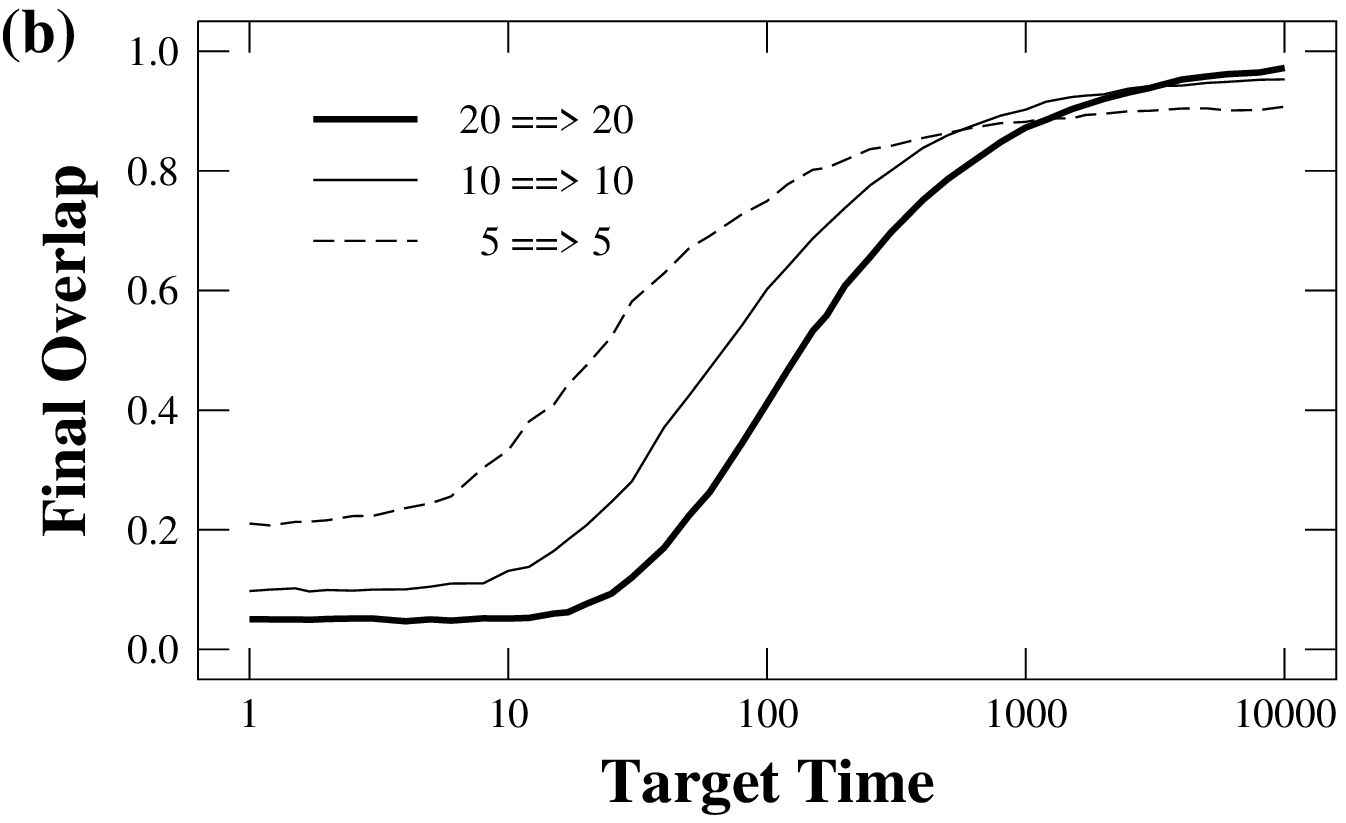}
\caption{Final overlaps by $\varepsilon_{\rm a}(t)$ for the transition
between generic states constructed from 5 (dashed curve),
10 (thin curve), and 20 eigenstates (thick curve)
in (a) GOE and (b) GUE random matrix systems.
Each curve is obtained as an average over 100 different random configurations
of $H_0$, $V$, $\{c_j\}$, and $\{d_j\}$.}
\label{fig:m2m}
\end{center}
\end{figure}

In case of multilevel-multilevel transitions,
the same procedure can be executed with a numerical integration of
the Schr\"odinger equation without the RWA.
In this case, control fields $\varepsilon_{\rm a}(t)$ and
$\varepsilon_{\rm cg}(t)$ is determined by
$H_0$, $V$, $\{c_j\}$, and $\{d_j\}$,
which are created from pseudo-random numbers
according to distribution functions under each universality class.
Fig.\ref{fig:m2m} shows the performance of $\varepsilon_{\rm a}(t)$
for GOE and GUE random matrices.
Almost the same properties can be observed as the case of the transition
from an eigenstate.
However, even if we take the limit $T\rightarrow\infty$,
the final overlap $|\langle\Phi_T|\psi(T)\rangle|$ does not
seem to converge to $1$.
This is because $\varepsilon_{\rm a}(t)$ is an approximate
field obtained by ignoring quantities with the order of $1/N$.

\begin{figure}
\begin{center}
\includegraphics[scale=0.55]{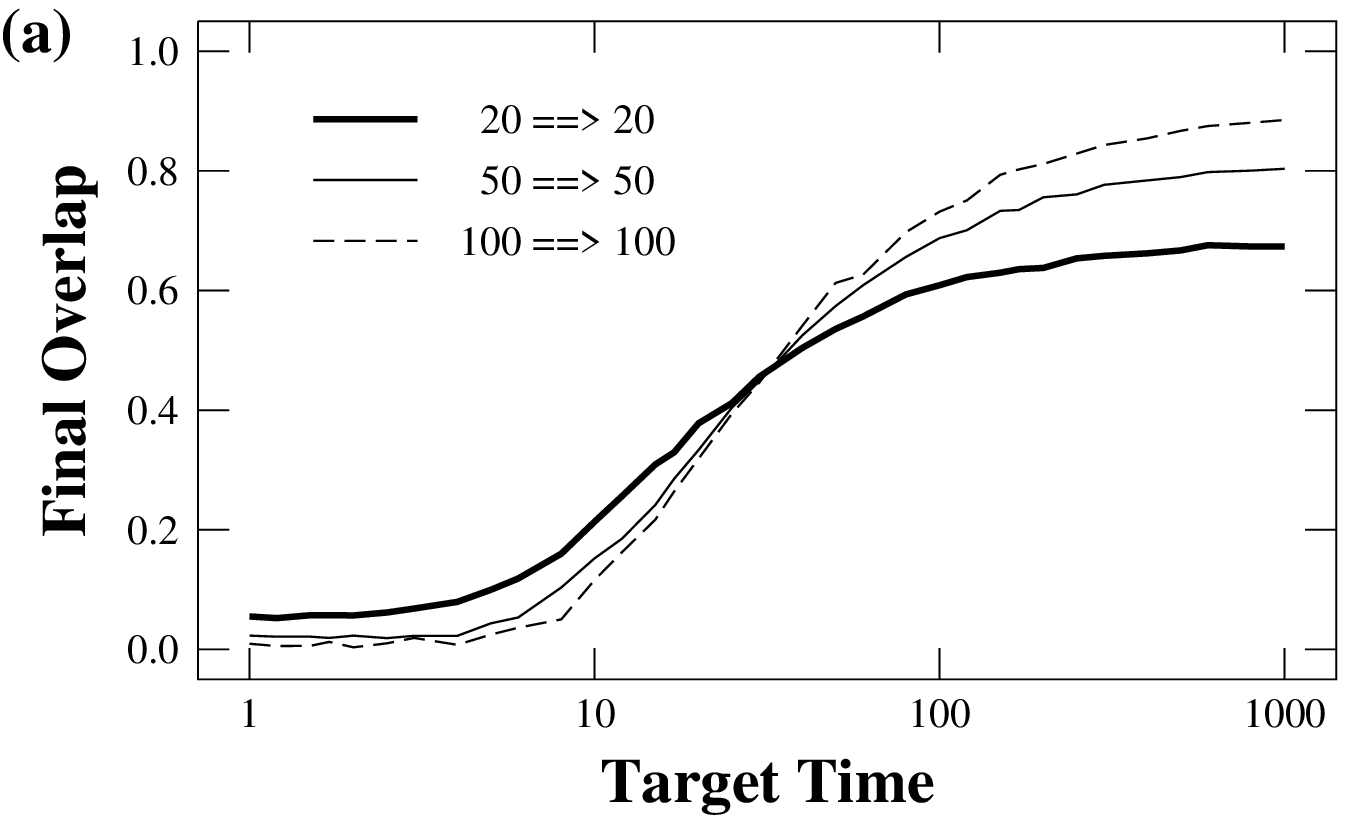}\\
\includegraphics[scale=0.55]{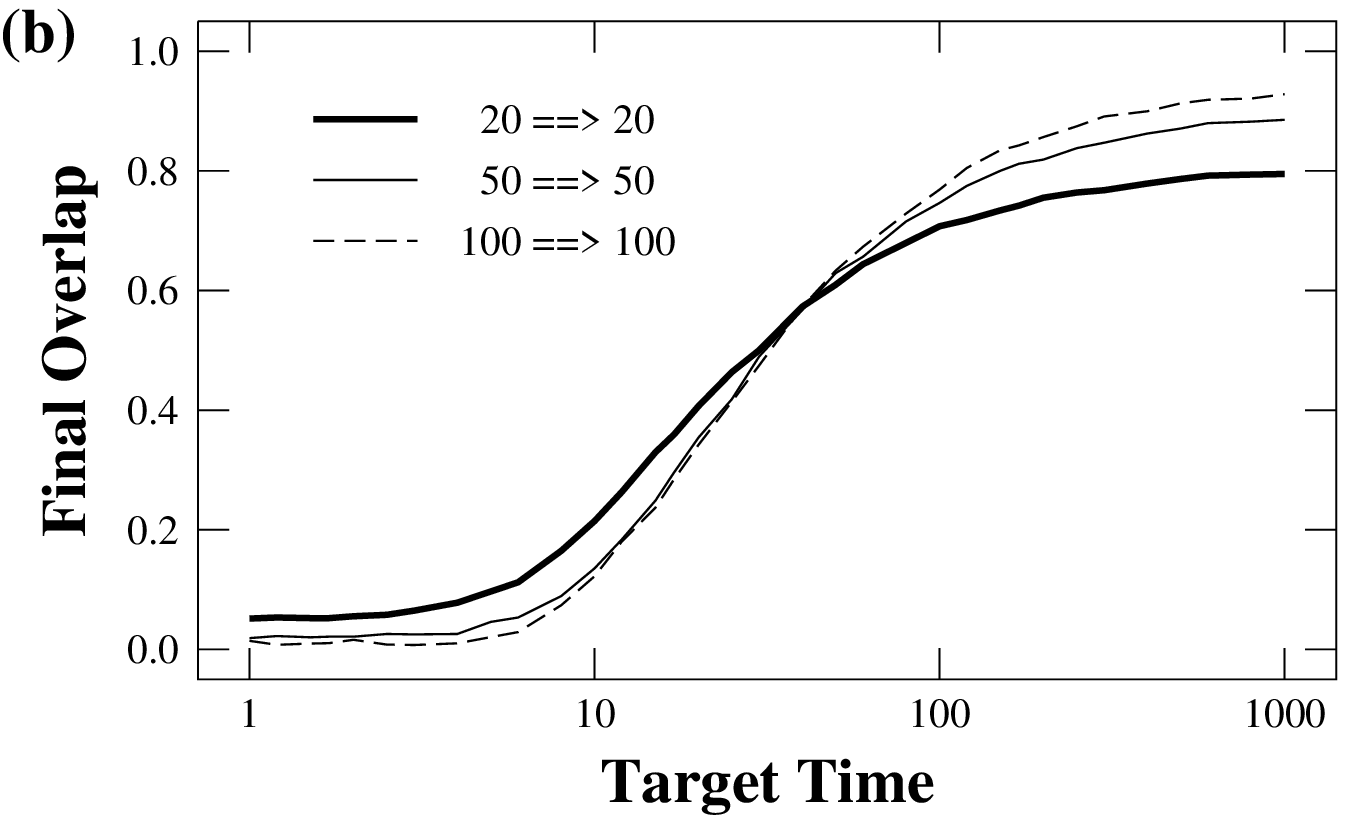}
\caption{Final overlaps by $\varepsilon_{\rm cg}(t)$ are shown
for the transition between generic states constructed
from 20 (thick curve), 50 (thin curve), and 100 eigenstates (dashed curve)
in (a) GOE and (b) GUE random matrix systems.
Each curve is obtained as an average over 100 different random configurations
of $H_0$, $V$, $\{c_j\}$, and $\{d_j\}$.}
\label{fig:cg}
\end{center}
\end{figure}

Fig.\ref{fig:cg} is the result by $\varepsilon_{\rm cg}(t)$.
In GOE case (Fig.\ref{fig:cg}(a)),
much improvement can be seen
when it is compared to one by $\varepsilon_{\rm a}(t)$,
while it is almost the same for GUE (Fig.\ref{fig:cg}(b)).
The overlap does not converge to 1
in the limit of $T\rightarrow\infty$ since
$\varepsilon_{\rm cg}(t)$ is valid under $N\rightarrow\infty$.
The important thing observed here is that,
in case of $\varepsilon_{\rm cg}(t)$,
longer target times are not necessary even for larger matrices.
This is significant in the practical application of controlling large systems.

\section{Extension to Realistic Chaotic Systems}

\begin{figure}
\begin{center}
\includegraphics[scale=0.18]{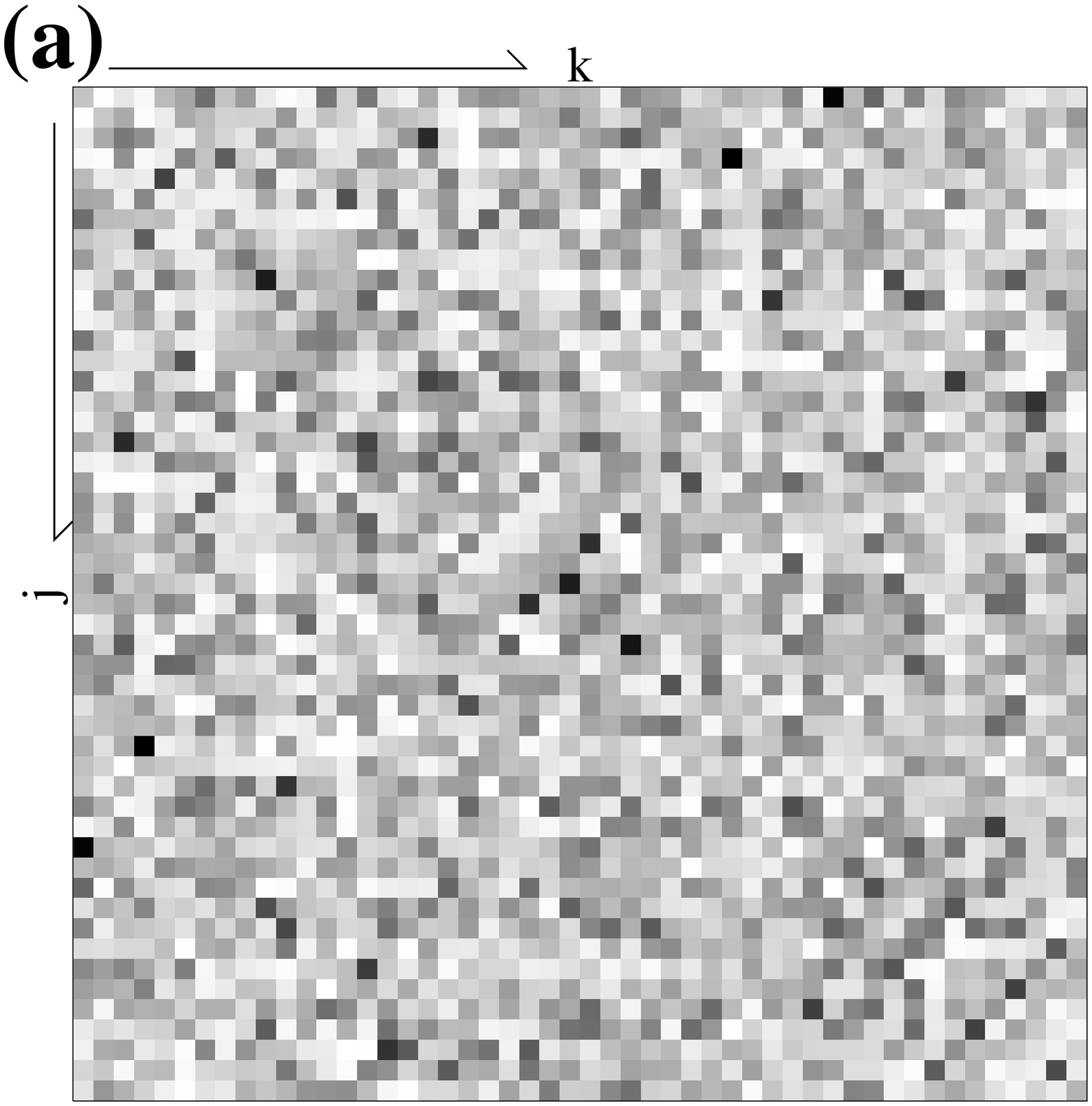}
\includegraphics[scale=0.18]{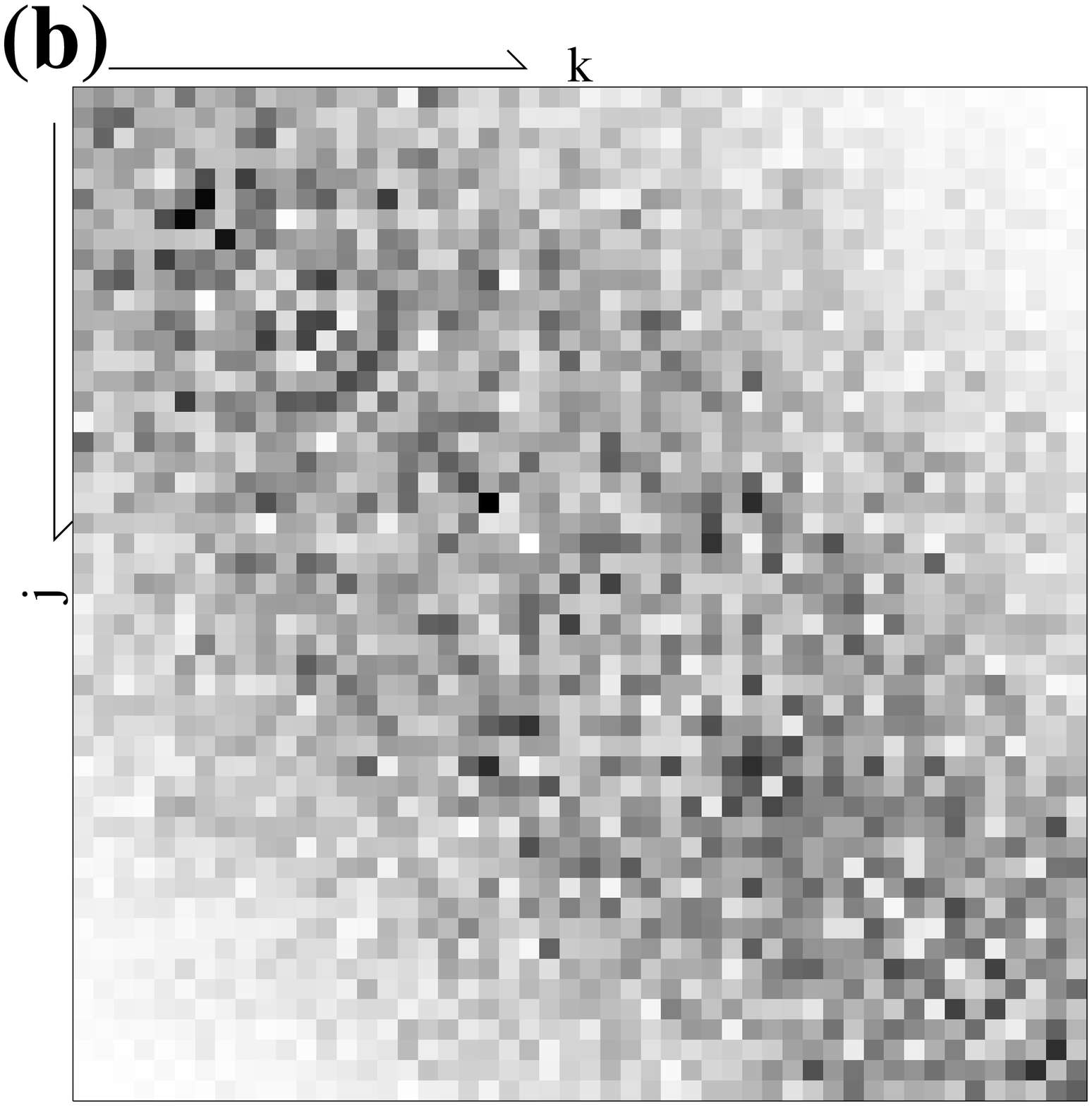}
\caption{
Examples of the interaction Hamiltonian V ($50\times50$).
The magnitude of matrix elements are shown for
(a) full random matrix subject to GOE, and
(b) a banded random matrix subject to GUE ($\Delta_0=15$).}
\label{fig:randomMatrix}
\end{center}
\end{figure}

The full random matrix systems studied in the previous section
is a model of strongly chaotic systems.
In order to consider realistic quantum systems,
one of candidates to be studied is a banded random matrix \cite{CIK00}.
In this section, we try to improve
our external field $\varepsilon_{\rm cg}(t)$
for the case of the banded random matrix \cite{TF07AIP}.

\subsection{Analytic field for banded random matrix systems}

We consider the case that the interaction Hamiltonian $\hat V$ is
a banded random matrix in the eigenstate representation of $\hat H_0$.
The elements of $V$ are random complex numbers with the following distribution
\begin{equation}
\label{eqn:V-brm}
  \left\langle\left|V_{jk}\right|^2\right\rangle\equiv\left\langle
    \left|\langle\phi_j|\hat V|\phi_k\rangle\right|^2
  \right\rangle
  =\exp\left[-\frac{(E_j-E_k)^2}{{\Delta_0}^2}\right].
\end{equation}
We introduce a new field $\varepsilon_{\rm brm}(t)$
as an extension of $\varepsilon_{\rm cg}(t)$,
\begin{equation}
  \varepsilon_{\rm brm}(t)=\sum_{jk}{\rm Re}\left[
    A_{jk}c_j^*V_{jk}d_k\ e^{iw_{jk}t}
  \right],
\end{equation}
with an extra-amplitude factor $A_{jk}$.
The coefficients $a_j(t)$ in Eq.(\ref{eq:es-rep})
satisfy the Schr\"odinger equation
\begin{equation}
\label{eqn:rwa-cg}
  i\hbar\frac{d}{dt}a_k(t)=\frac12\sum_j\left[
    A_{jk}c_j^*d_k+A_{kj}^*c_kd_j^*
  \right]\left|V_{jk}\right|^2a_j(t)
\end{equation}
after ignoring all the oscillating terms.
If we assume that the transition is smooth,
$a_j(t)$ should be written as 
\begin{equation}
  a_k(t)=c_k\cos\left(\frac{\pi t}{2T}\right)
        -id_k\sin\left(\frac{\pi t}{2T}\right),
\end{equation}
where we take $\Theta=0$ in Eq.(\ref{eq:innerProduct}) for simplicity.
Substituting these coefficients into Eq.(\ref{eqn:rwa-cg}),
we obtain a relation
\begin{eqnarray}
  &&\frac{i\pi\hbar}{2T}\left[
    -c_k\sin\left(\frac{\pi t}{2T}\right)
    -id_k\cos\left(\frac{\pi t}{2T}\right)
  \right]\nonumber \\
  &&=\frac12\sum_j\left[
      A_{jk}\left|c_j\right|^2d_k\cos\left(\frac{\pi t}{2T}\right)
    \right.\nonumber\\
    &&\qquad\qquad\left.
     -iA_{kj}^*c_k\left|d_j\right|^2\sin\left(\frac{\pi t}{2T}\right)
  \right]\left|V_{jk}\right|^2
\end{eqnarray}
under the assumption of random phases \cite{TF07}.
Finally, we obtain the conditions for the coefficients $A_{jk}$
\begin{equation}
  A_{jk}=A_{kj}^*\simeq\frac{\pi\hbar}{T}\exp\left[
    \frac{(E_j-E_k)^2}{{\Delta_0}^2}
  \right].
\end{equation}
If we consider the case that
those coefficients $c_j$ and $d_k$ have Gaussian distribution functions
in the energy space,
\begin{eqnarray}
  \langle|c_j|^2\rangle
    &\propto&\exp\left[-\frac{(E_j-E_c)^2}{{\Delta_c}^2}\right],\nonumber\\
  \langle|d_k|^2\rangle
    &\propto&\exp\left[-\frac{(E_k-E_d)^2}{{\Delta_d}^2}\right],
\label{eqn:initial_target-brm}
\end{eqnarray}
with centers $E_c$ and $E_d$ and widths $\Delta_c$ and $\Delta_d$,
we can define
\begin{equation}
\label{eqn:field-brm}
  \varepsilon_{\rm brm}(t)=\frac{\pi\hbar}{T}\sum_{jk}{\rm Re}\left[
    c_j^*V_{jk}d_k\ e^{iw_{jk}t}
  \right]\exp\left[\frac{(E_j-E_k)^2}{{\Delta_0}^2}\right].
\end{equation}
This field has a finite amplitude only when 
\begin{equation}
\Delta_c<\Delta_0\quad\hbox{and}\quad\Delta_d<\Delta_0.
\end{equation}
If not, the field has an infinite amplitude
in the limit of $E_j, E_k\rightarrow\pm\infty$
due to the exponential factor $A_{jk}$ in Eq.(\ref{eqn:field-brm}).
Thus, the analytic field is refined
when the widths of the initial and target states are relatively small
compared to the width of the banded random matrix elements.
\eject

\subsection{Numerical evaluation}

\begin{figure}
\begin{center}
\includegraphics[scale=0.55]{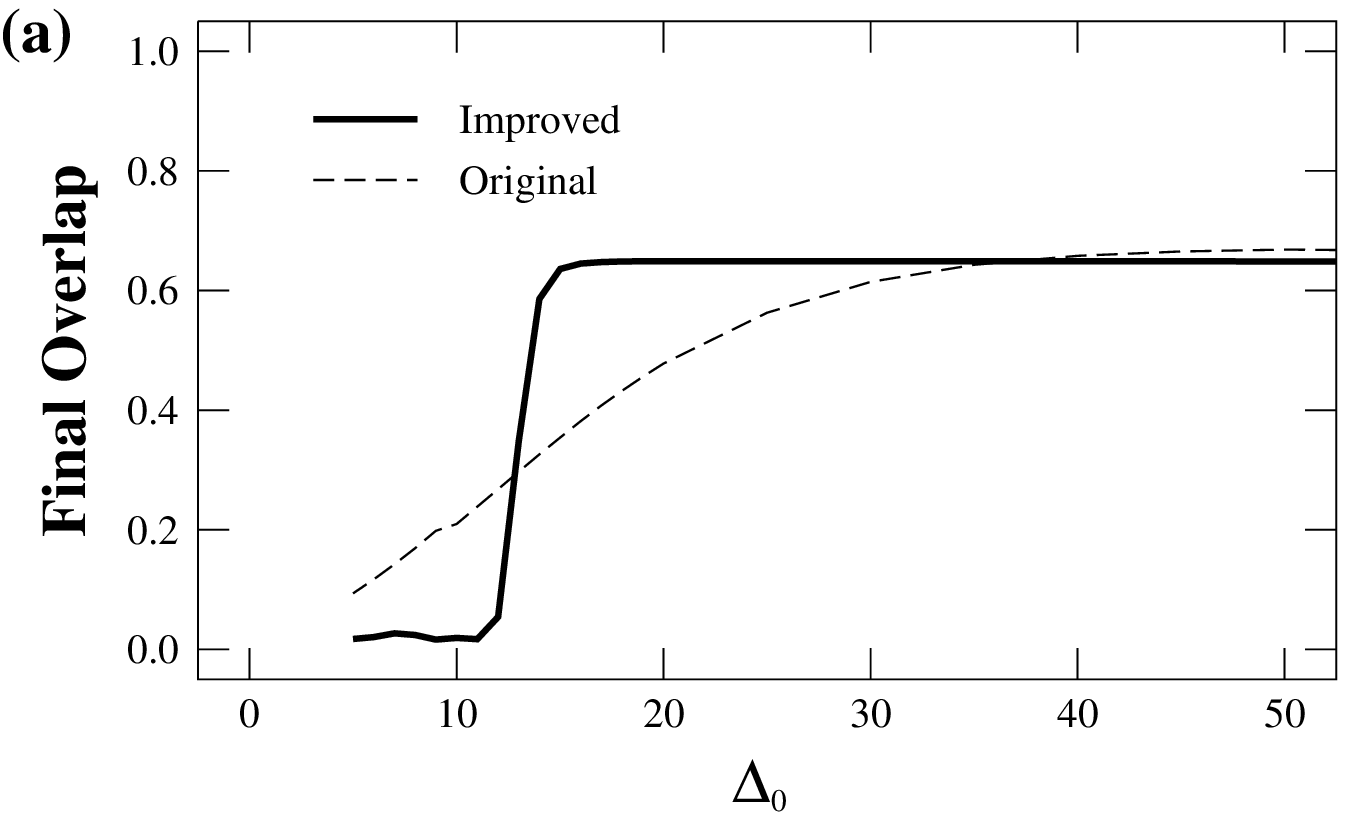}\\
\includegraphics[scale=0.55]{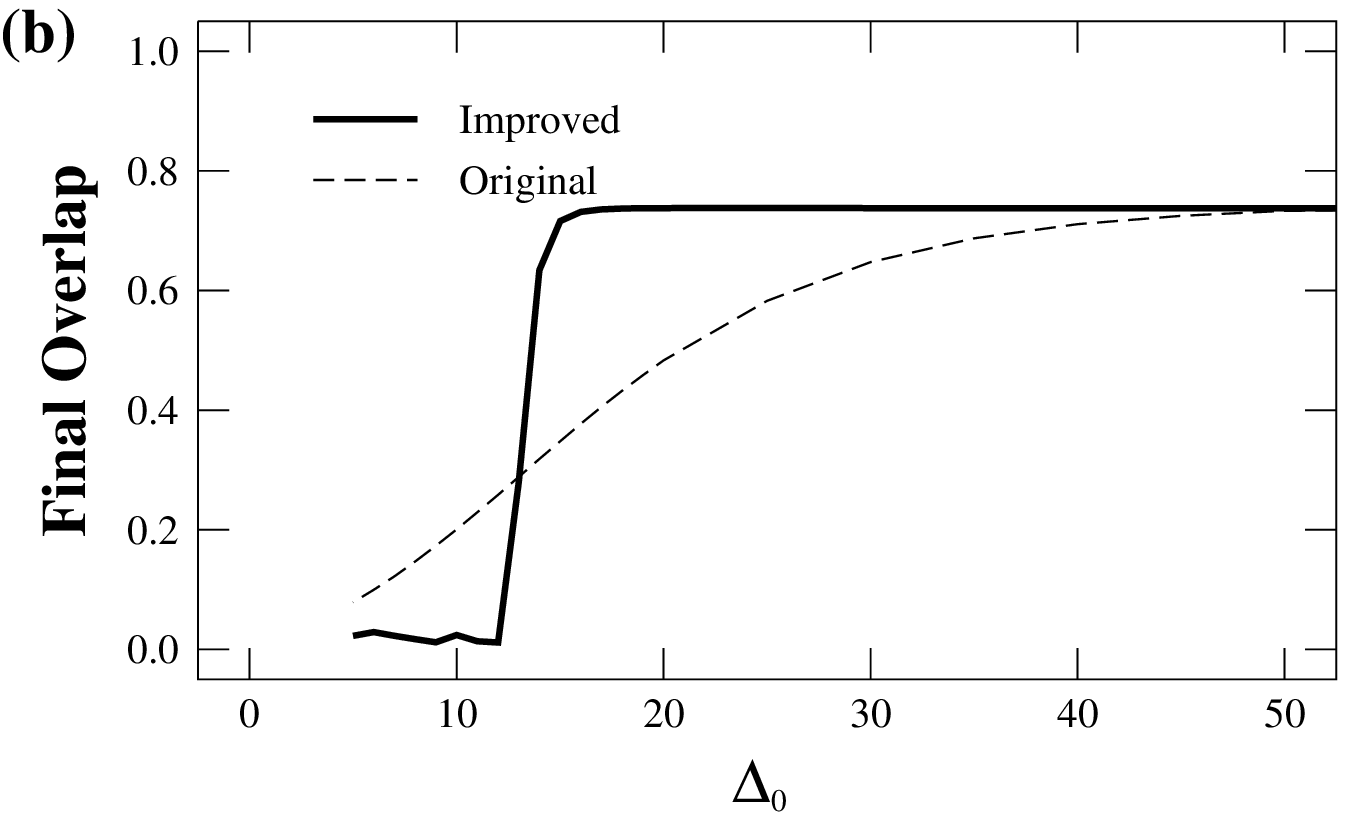}
\caption{Improvements by $\varepsilon_{\rm brm}(t)$ is shown for
(a) GOE case and (b) GUE case.
The initial and the target states are both Gaussian wavepacket
in the energy space with $\Delta_c=\Delta_d=15$.
}
\label{fig:result}
\end{center}
\end{figure}

\begin{figure}
\begin{center}
\includegraphics[scale=0.55]{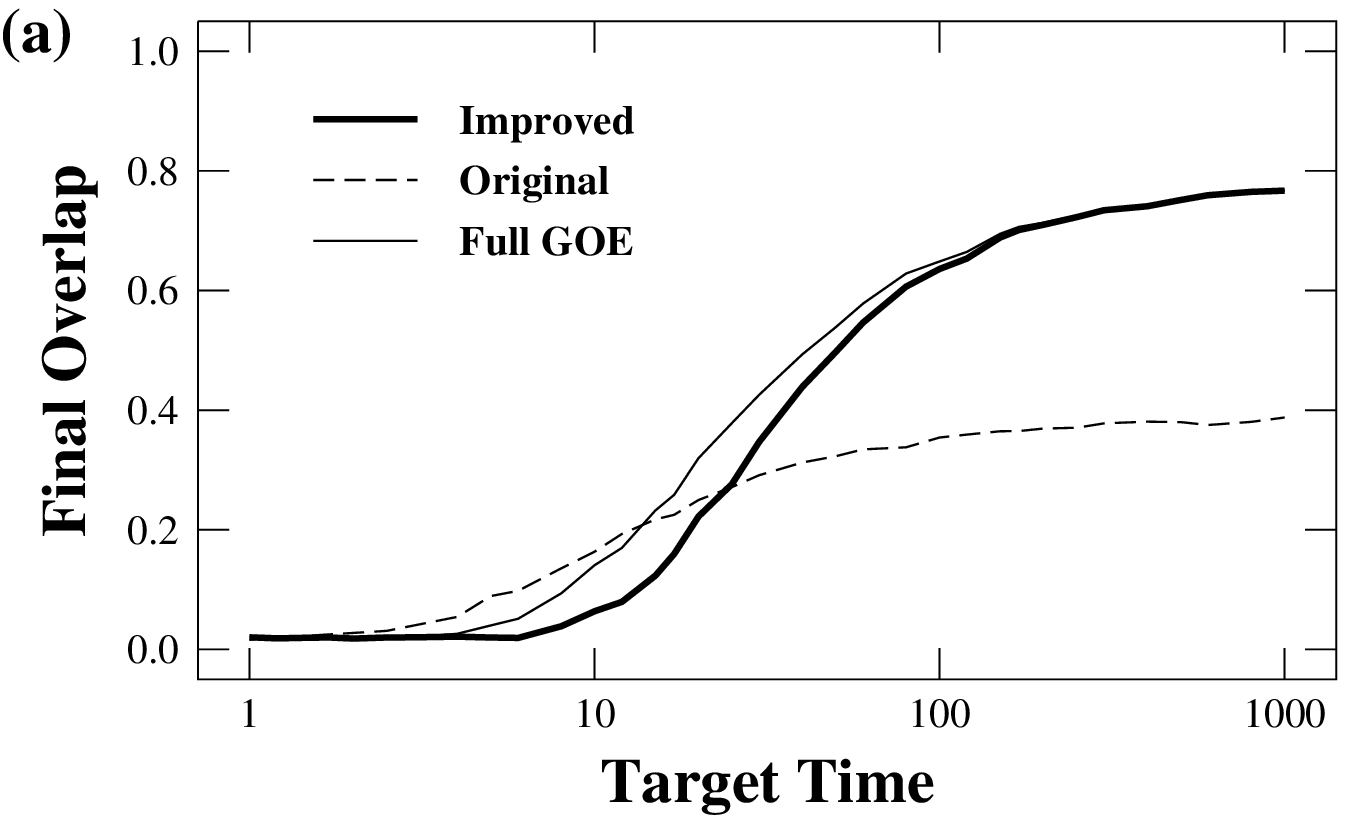}\\
\includegraphics[scale=0.55]{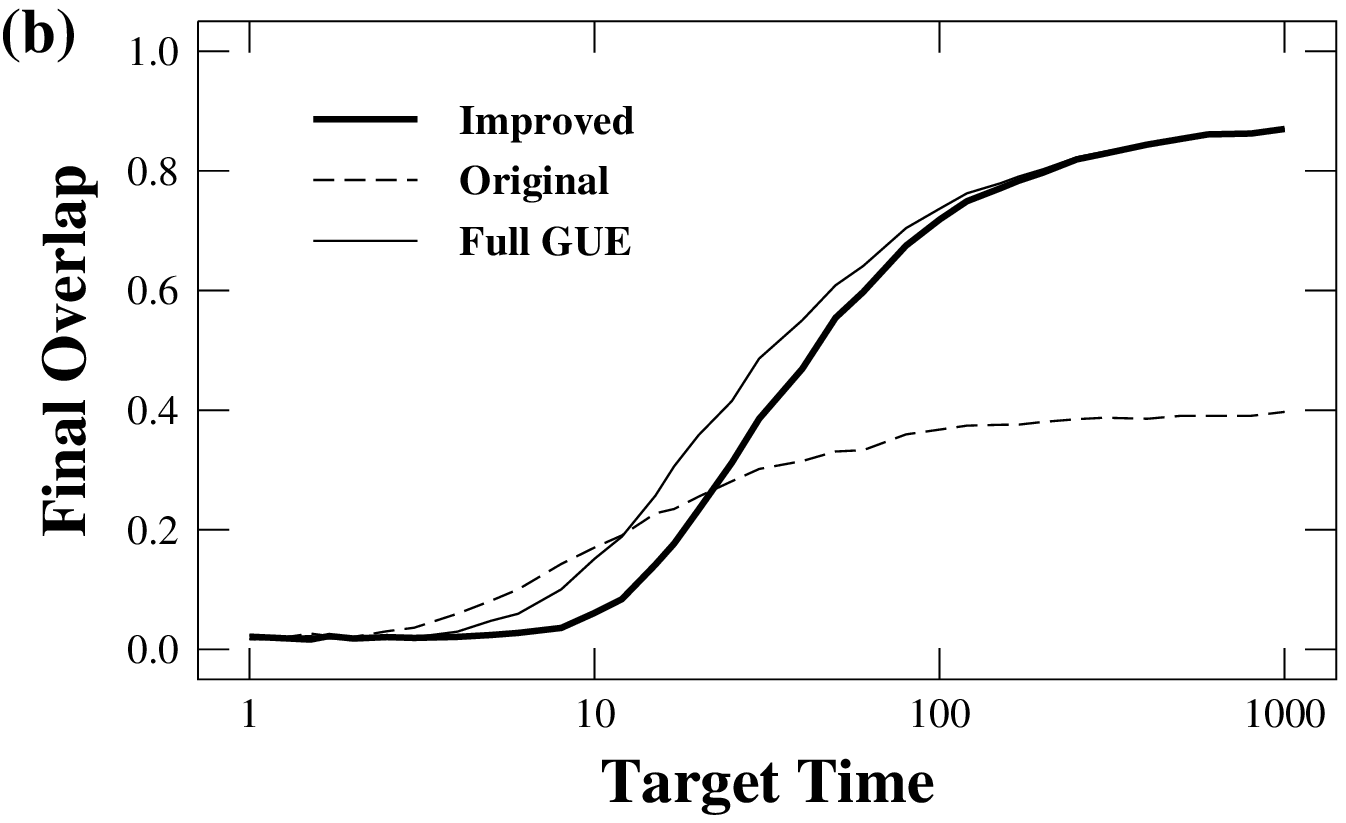}
\caption{Improved final overlap by $\varepsilon_{\rm brm}(t)$ (thick curve)
is compared to that by $\varepsilon_{\rm cg}(t)$ (dashed curve)
for the case of banded random matrix systems with
$\Delta_c=\Delta_d=\Delta_0=15$ subject to (a) GOE and (b) GUE.
The result by $\varepsilon_{\rm cg}(t)$ in a full random matrix system
is also shown for comparison (thin curve).
Each curve is obtained as an average over 100 different random configurations
of $H_0$, $V$, $\{c_j\}$, and $\{d_j\}$.}
\label{fig:band}
\end{center}
\end{figure}

We confirm the validity of $\varepsilon_{\rm brm}(t)$
for the system with a banded random-matrix interaction.
The numerical test is configured as follows.
The initial and target states are defined
as quantum vectors with random complex coefficients $c_j$ and $d_j$
subject to Eq.(\ref{eqn:initial_target-brm}).
Here, we choose $\Delta_c=\Delta_d=15$, $E_c=-10$, and $E_d=10$,
where $H_0$ is a $50\times50$ random matrix of GOE and GUE,
and is scaled so that
its eigenvalues $\{E_j\}$ are distributed in an interval $[-25,\ 25]$.
The interaction Hamiltonian $\hat V$ is also a $50\times50$ matrix
while its elements obey a banded-random distribution
in the eigenstate representation of $\hat H_0$ with $\Delta_0$.

The optimal field Eq.(\ref{eqn:field-brm}) is calculated
from those quantities $\{c_j\}$, $\{d_k\}$, $\{V_{jk} \}$, and $\{E_j\}$
with parameters $T$ and $\Delta_0$.
In order to check the validity of $\varepsilon_{\rm brm}(t)$
Eq.(\ref{eqn:field-brm}),
we solve the initial value problems with Hamiltonian Eq.(\ref{eq:Hamiltonian})
driven by Eq.(\ref{eqn:field-brm})
for various band widths $\Delta_0$ of the interaction Hamiltonian $\hat V$.
The results are shown in Figure \ref{fig:result}.
When we use the original analytic field $\varepsilon_{\rm cg}(t)$
Eq.(\ref{eq:CG-field}),
the performance of the optimal field (dashed curve) decreases
for the banded matrices with smaller widths.
On the other hand, the final overlaps (solid curve) by the refined
analytic field Eq.(\ref{eqn:field-brm}) does not change
even for the smaller width
untill the limit $\Delta_0\approx\Delta_c$ or $\Delta_0\approx\Delta_d$.
In Fig.\ref{fig:band}, the performance of $\varepsilon_{\rm brm}(t)$ is
shown for the various target time $T$.
The final overlap by $\varepsilon_{\rm brm}(t)$ (thick curve)
is much improved from the result by $\varepsilon_{\rm cg}(t)$ (dashed curve),
which is comparable to the level of the original result
by a full random matrix (thin curve).

\section{Conclusion}

We studied several forms of analytic fields
to steer quantum states in generic systems.
Through numerical evaluations of those fields,
the rotating-wave approximation for multi-level systems
are also validated in the limit of the large target time.
It was shown that the analytic field defined through
the coarse-grained idea can be used in the limit of many states.
We extended our previous analytic approaches for controlling complex
quantum systems to deal with more realistic systems with a banded
random matrix.  The key ingredient is the amplitude factor $A_{jk}$,
which is an exponentially growing function, introduced in the analytic 
optimal field Eq.(\ref{eqn:field-brm}).  We showed that the new 
analytic field outperforms the previous field Eq.(\ref{eq:CG-field})
for a full random matrix.
In the near future we will apply this optimal field
to quantum chaos systems such as quantum kicked rotors (top)
\cite{FMT03} and to more realistic molecular systems \cite{FYHS07,FZS06}.
\eject

\begin{widetext}
\appendix
\section{Derivation of the Exact Field $\varepsilon_{\rm e}(t)$}
\label{apx:E-field}

Substitution of the restricted form Eq.(\ref{eq:Er-field})
into Eqs.(\ref{eq:s2m1}) and (\ref{eq:s2m2}) gives relations
\begin{equation}
  \dot\theta=-\frac{1}{i\hbar}\sum_{j'\ne0}\left(
    \varepsilon_{j'0}\ e^{-i\omega_{j'0}(t-T)}+{\rm c.c.}
  \right)\left[
    V_{00}\cot\theta+\sum_{j\ne0}V_{0j}d_j\ e^{-i\omega_{j0}(t-T)}
  \right]
\end{equation}
and
\begin{equation}
  \dot\theta\ d_j=\frac{1}{i\hbar}\sum_{j'\ne0}\left(
    \varepsilon_{j'0}\ e^{-i\omega_{j'0}(t-T)}+{\rm c.c.}
  \right)\left[
    V_{j0}\ e^{i\omega_{j0}(t-T)}
    +\sum_{k\ne0}V_{jk}d_k\ e^{i\omega_{jk}(t-T)}\tan\theta
  \right],
\end{equation}
\end{widetext}
respectively.
By dropping all the oscillating terms from these expressions,
we obtain
\begin{equation}
  i\hbar\ \dot\theta=-\sum_{j\ne0}V_{0j}d_j\ \varepsilon_{j0}^*,\qquad
  i\hbar\ d_j\dot\theta=V_{j0}\ \varepsilon_{j0}
\end{equation}
as relations under RWA.
From the second relation, $\varepsilon_{j0}$ is determined to be
\begin{equation}
\label{eq:E-amp}
  \varepsilon_{j0}=\frac{i d_j}{V_{j0}}\hbar\Omega,
\end{equation}
and it is easily shown that this definition also satisfies the first
relation if we note $V_{0j}=V_{j0}^*$ and $\sum|d_j|^2=1$.
Thus, the exact control field under RWA is obtained as
Eq.(\ref{eq:E-field}).

\section{Derivation of the Approximate Field $\varepsilon_{\rm a}(t)$}
\label{apx:A-field}

By substituting Eqs.(\ref{eq:state-phi}) and (\ref{eq:state-chi}) into
Eqs.(\ref{eq:theta-dot1}) and (\ref{eq:theta-dot2}),
we obtain equations for $\varepsilon(t)$ in the eigenstate representation,
\begin{eqnarray}
  \dot\theta&=&-\frac{\varepsilon(t)}{i\hbar}\sum_{j,k}c_j^*\left(
    c_k\cot\theta-ie^{-i\alpha}\tilde d_k
  \right)V_{jk}\ e^{i\omega_{jk}t},\\
  \dot\theta&=&\frac{\varepsilon(t)}{i\hbar}\sum_{j,k}\tilde d_j^*\left(
    ie^{i\alpha}c_k+\tilde d_k\tan\theta
  \right)V_{jk}\ e^{i\omega_{jk}t},
\end{eqnarray}
respectively.
If we assume the restricted form Eq.(\ref{eq:field-restriction})
as the control field, these equations can be simplified to
\begin{equation}
\label{eq:theta-dot1-rwa}
  \dot\theta=-\frac{1}{i\hbar}\sum_{j,k (\ne j)}
    \left(\varepsilon_{jk}+\varepsilon_{kj}^*\right)c_j^*\left(
      c_k\cot\theta-ie^{-i\alpha}\tilde d_k
    \right)V_{jk},
\end{equation}
\begin{equation}
\label{eq:theta-dot2-rwa}
  \dot\theta=\frac{1}{i\hbar}\sum_{j,k (\ne j)}
    \left(\varepsilon_{jk}+\varepsilon_{kj}^*\right)\tilde d_j^*\left(
      ie^{i\alpha}c_k+\tilde d_k\tan\theta
    \right)V_{jk},
\end{equation}
after dropping all the oscillating terms.
The problem is to determine the amplitude factors $\varepsilon_{jk}$
which satisfy these equations for given $\dot\theta$.

Here, we try to define a simple solution that is independent of
$\theta$.  Such a solution, if exist, should satisfies
Eqs.(\ref{eq:theta-dot1-rwa}) and (\ref{eq:theta-dot2-rwa})
for any value of $\theta$.  From Eq.(\ref{eq:theta-dot1-rwa})
with $\theta=\pi/2$ or Eq.(\ref{eq:theta-dot2-rwa}) with $\theta=0$,
a condition for $\varepsilon_{jk}$,
\begin{equation}
\label{eq:required}
  \dot\theta=\frac{e^{-i\alpha}}{\hbar}\sum_{j,k (\ne j)}
    \left(\varepsilon_{jk}+\varepsilon_{kj}^*\right)
    c_j^*V_{jk}\tilde d_k
\end{equation}
is required.
While it is still difficult to define $\varepsilon_{jk}$
satisfying this equation exactly,
an approximate one is defined when the number of states $N$ is large enough.
By introducing amplitudes,
\begin{equation}
\label{eq:A-amp}
  \varepsilon_{jk}=\frac{e^{i\alpha}c_j\tilde d_k^*}{V_{jk}}\hbar\Omega,
\end{equation}
as an extension of Eq.(\ref{eq:E-amp}),
the required Eq.(\ref{eq:required})
leads to Eq.(\ref{eq:theta-dot}) by the use of
\begin{eqnarray}
  \sum_{j,k(\ne j)}c_j^*c_j\tilde d_k^*\tilde d_k&=&1+O(N^{-1}),\nonumber\\
  \sum_{j,k(\ne j)}c_j^*\tilde d_jc_k^*\tilde d_k&=&O(N^{-1}),
\end{eqnarray}
under the assumption
that there are no special correlations between $c_j$ and $\tilde d_j$
other than the orthogonal relation $\sum_jc_j^*\tilde d_j=0$.

Moreover, Eq.(\ref{eq:A-amp}) approximately satisfies
Eqs.(\ref{eq:theta-dot1-rwa}) and (\ref{eq:theta-dot2-rwa})
for any $\theta$, because small quantities
\begin{equation}
  \sum_{j,k(\ne j)}c_j^*c_j\tilde d_k^*c_k,\quad
  \sum_{j,k(\ne j)}\tilde d_j^*\tilde d_jc_k^*\tilde d_k,
\end{equation}
of the order $O(N^{-1})$ are also ignored in the limit of $N\rightarrow\infty$.
Thus, the control field Eq.(\ref{eq:A-field}) is approximately valid
for a transition between multi-level states.

\bibliography{BibTeX/General,BibTeX/QuantumChaos,BibTeX/Control,BibTeX/QuantumDynamics,BibTeX/MyWorks}

\end{document}